\documentclass[twocolumn,showpacs,preprintnumbers,amsmath,amssymb]{revtex4}

\usepackage{graphicx}
\usepackage{dcolumn}
\usepackage{bm}
\usepackage{color}

\begin{document}


\title{Synchronization on Effective Networks}

\author{Tao Zhou$^{1,2}$}
\author{Ming Zhao$^{1,3,4}$}
\author{Changsong Zhou$^3$}
\email{cszhou@hkbu.edu.hk}
\affiliation{$1$Department of Modern
Physics, University of Science and
Technology of China, Hefei 230026, PR China\\
$^2$Department of Physics, University of Fribourg, Chemin du Mus\'ee
3, CH-1700 Fribourg, Switzerland \\
$^3$Department of Physics,
Centre for Nonlinear Studies, and The Beijing-Hong Kong-Singapore
Joint Centre for Nonlinear and Complex Systems (Hong Kong), Hong
Kong Baptist University, Kowloon Tong, Hong Kong, China\\
$^4$College of Physics and Technology, Guangxi Normal University,
Guilin 541004, P. R. China}

\begin{abstract}

The study of network synchronization has attracted increasing
attention recently. In this paper, we strictly define a class of
networks, namely effective networks, which are synchronizable and
orientable networks. We can prove that all the effective networks
with the same size have the same spectra, and are of the best
synchronizability according to the master stability analysis.
However, it is found that  the synchronization time for different
effective networks can be quite different. Further analysis show that
the key ingredient affecting the synchronization time is the
maximal depth of an effective network: the larger depth results in
a longer synchronization time. The secondary factor is the number
of links. The more links connecting the nodes in the same layer
(horizontal links) will lead to longer synchronization time, while
the increasing number of links connecting nodes in neighboring
layers (vertical links) will accelerate the synchronization. Our
findings provide insights into the roles of horizontal and vertical
links in synchronizing process, and suggest that the spectral
analysis is helpful yet insufficient for the understanding of
network synchronization.

\end{abstract}


\maketitle

Synchronization is observed in many natural, social, physical and
biological systems, and has found applications in a variety of
fields \cite{Strogatz2003}. As a result, the subject of
synchronization is continuously calling for serious and systematic
investigation, and has evolved to be an independent field of
scientific research. Recently, the dramatically increasing
interests in complex networks have been pervading the study of
synchronization---understanding the synchronizing process of a
network of dynamical systems is attracting more and more attention
(see the review articles \cite{Arenas,Chen} and the references
therein). A particularly interesting issue is how to enhance the
network synchronizability \cite{Zhao2007}. Under the framework of
master stability analysis \cite{Pecora1998,Barahona2002}, if the
synchronization region is bounded and the nodes are coupled
linearly and symmetrically, the network synchronizability can be
measured by the eigenratio, which is defined as the ratio of the
largest eigenvalue to the smallest nonzero eigenvalue of the
coupling matrix. A smaller eigenratio indicates a better
synchronizability, and vice versa. Some effective
synchronizability-enhancement methods, aiming at reducing the
eigenratio, are proposed. These methods include the discrete
optimization of network structure \cite{Donetti2005,Wang2007}, the
regulation of coupling pattern
\cite{Motter2005,Motter2005b,Hwang2005,Chavez2005,Zhao2006,Lu2007},
the adaptive coupling strategy \cite{Zhou2006,Huang2006} and the
structural modification \cite{Zhao2005,Zhou2005,Yin2006}. It is
worth noting, although the coupling matrix of coupling pattern
regulated networks and adaptive networks may become asymmetric,
which correspond to directed networks, the master stability
analysis is still valid \cite{Pecora1998}.


Recently, Nishikawa and Motter \cite{Nishikawa2006,Nishikawa2007}
proposed a class of maximally synchronizable networks that have
the smallest eigenratio, $R=1$. They showed that a network is
maximally synchronizable if (i) it embeds an oriented spanning
tree, (ii) it does not have any directed loops, and (iii) it has
normalized input coupling strength of each node. Their work
\cite{Nishikawa2006,Nishikawa2007} provides a start point to the
understanding of the role of directivity and feedback loops in
network-based synchronization \cite{Um2008}, and is also relevant
to the synchronization control \cite{Wang2002,Sorrentino2007}.
However, some crucial issues, such as the time required for full
synchronization, lack serious consideration thus far.

In this paper, using the language of graph theory
\cite{Bollobas1998}, we define a new class of networks, namely
\emph{effective networks}, as the \emph{synchronizable} and
\emph{orientable} networks, and prove that the coupling matrix of
any effective network has eigenvalues
$0=\lambda_1<\lambda_2=\cdots \lambda_N=1$. We further propose a
preferential selection algorithm (PSA) to extract effective
network from any given connected undirected network, and
investigate the synchronization of chaotic Logistic systems
\cite{May1976} on effective networks, emphasizing on the relation
between the topology and synchronization time. Surprisingly,
although the spectra of all effective networks are exactly the
same, there is a great difference between their synchronization
time. Based on extensive simulations, it is found that the
synchronization time is mainly determined by the maximal depth of
an effective network, complementary to a recent study about the
synchronization of the Kuramoto model in an ideal tree
\cite{Zeng2009}. The secondary factor affecting the
synchronization time is the number of links. The more the links
connecting the nodes in the same layer (\emph{horizontal links}),
the longer the synchronization time. Our simulation suggests a
logarithmic relation between the synchronization time and the
number of those links. In contrast, the increasing number of links
connecting nodes in neighboring layers (\emph{vertical links})
will accelerate the synchronization speed. Given the maximal
depth, the simulation results indicate that when the network
consists of abundant horizontal and vertical links, the ratio of
the number of horizontal links to the number of vertical links
mainly determines the synchronization time.

\begin{figure}
\scalebox{0.65}[0.65]{\includegraphics{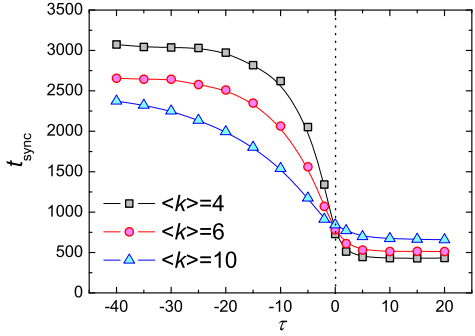}}
\caption{\label{fig:epsart} (Color online) $t_{sync}$ vs. $\tau$ for
different average degrees, $\langle k\rangle$=4, 6, and 10. The dot
line indicates the crossover. The network size is 1000, and the
coupling strength takes 0.55. Each data point is obtained by
averaging over 100 independent realizations.}
\end{figure}

We denote $D(V,E)$ a directed graph, where $V$ is the set of node
and $E$ the set of directed links. The multiple links and
self-connections are not allowed. A directed network is called
\emph{synchronizable} if for any two different nodes $v_i$ and
$v_j$, $\exists P(v_i\rightarrow v_j)$, or $\exists
P(v_j\rightarrow v_i)$, or $\exists k\neq i,j$ such that $\exists
P(v_k\rightarrow v_i)$ and $\exists P(v_k\rightarrow v_j)$.
Here, $P(v_i\rightarrow v_j)$ denotes an arbitrary directed path from
$v_i$ to $v_j$. In language, any pair of nodes are either coupled directly or driven
by some common nodes.
A directed network is \emph{orientable} if it
contains no directed cycles. Note that, a synchronizable network
must be connected, while an orientable network is not necessarily
connected. We normalize the input coupling strength of each node:
$l_{ii}=1$, $l_{ij}=-1/k_i^-$ if the directed link from $v_j$ to
$v_i$ exists, otherwise $l_{ij}=0$. Here, $k_i^-$ denotes the
in-degree of node $v_i$. A node is called \emph{source} if its
in-degree equals zero, accordingly, we have a \emph{Lemma}: Any
orientable network has at least one source. Proof.---Assume there
exists an orientable network $D$ containing no source. Denote
$v_1\rightarrow v_2\rightarrow \cdots \rightarrow v_l$ the longest
directed path in $D$. Since $k_{v_1}^->0$, there exists at least a
link $v\rightarrow v_1$. For $D$ contains no directed loops,
$v\neq v_2, \cdots, v_l$, thus the directed path $v \rightarrow
v_1\rightarrow \cdots \rightarrow v_l$ is even longer, which
conflicts to the above assumption. Therefore, $D$ contains at
least one source. For effective network (i.e., synchronizable and
orientable network), we have a \emph{Theorem}: The normalized
Laplacian of an effective network, $L$, has eigenvalues
$0=\lambda_1<\lambda_2=\cdots \lambda_N=1$. Proof.---There exists
an ordered list of $V$ such that every link in $E$ is from a
former node to a later node. One can get this list from an empty
list according to the following steps: (i) Add all sources of the
current network at the top of the list in an arbitrary order; (ii)
Remove those source from the current network and go to step (i)
until the network becomes empty. This process is practicable since
the network is finite, and after each removal of sources, the
remain network is still orientable, thus has at least one new
source (see the \emph{Lemma}). Denote such an ordered list as
$Q=\{v_1,v_2,\cdots,v_N\}$, clearly, if $i>j$, the directed link
($v_i\rightarrow v_j$) can not exist because in the induced
subgraph of nodes $\{v_j,v_{j+1},\cdots,v_N\}$, $v_j$ is a source
without any in-link. Shift columns and rows of $L$ corresponding
to the ordered list $Q$ and denote it by $L_Q$, then, $L_Q$ is an
down-triangle matrix having the same spectrum as $L$. Therefore,
all the eigenvalues of $L_Q$, determined by the equation
$\texttt{det}(\lambda I-L_Q)=0$, are its diagonal elements. Any
node having in-degree larger than 0 corresponds to a unit diagonal
element, while a source corresponds to a zero diagonal element.
Clearly, $v_1$ is a source, and if there exists another source
$v_k$, since there are no directed paths end at $v_1$ or $v_k$,
$D$ is not synchronizable, conflicting to the condition that $D$
is an effective network. Actually, any effective network has one
and only one source, so the normalized coupling matrix has one
zero-eigenvalue and $N-1$ eigenvalues all equal 1.

\begin{figure}
\scalebox{0.85}[0.85]{\includegraphics{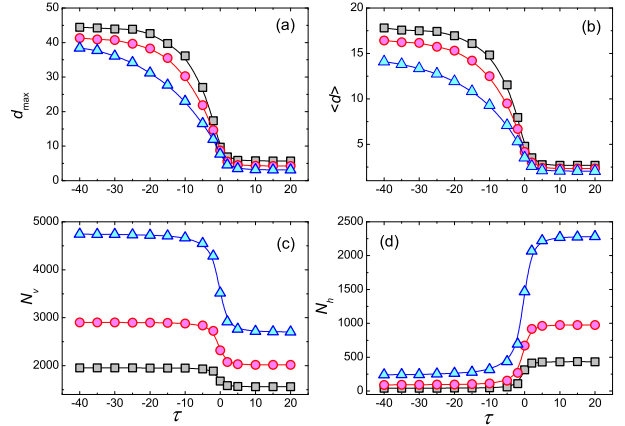}}
\caption{\label{fig:epsart} (Color online) The change of (a) the
maximal depth, (b) the average depth, (c) the number of vertical
links, $N_v$, and (d) the number of horizontal links, $N_h$, with
the parameter $\tau$ for average degree $\langle k\rangle$=4, 6,
and 10. The network is a BA scale-free network with $N$=1000. Each
data point is obtained by averaging over 100 independent
realizations. Symbols are of the same meanings as those in Fig.
1.}
\end{figure}

We propose an algorithm, namely \emph{preferential selection
algorithm} (PSA), to extract effective networks from any given
connected undirected network $G(V,E)$. The procedure is as
follows: (i) Set an empty list $Q$; (ii) Select a node $v_i$ in
$G$ with probability proportional to $k_i^\tau$ where $k_i$ is the
degree of $v_i$ and $\tau$ is a free parameter. Put this node at
the top of $Q$; (iii) Pick a node $v_j$ with probability
proportional to $k_j^\tau$, satisfying that it does not belong to
$Q$ and at least one of its neighbors belongs to $Q$ already. Push
it at the end of $Q$; (iv) Repeat the step (iii) until all nodes
in $V$ belong to $Q$. Since $G$ is connected, the termination
condition can always be achieved. Denoting $Q=\{v_1,v_2,...,v_N\}$
where $N$ is the size of $G$, then for each link $(v_i,v_j)\in E$,
we set its direction as $v_i\rightarrow v_j$ if $i<j$, or as
$v_j\rightarrow v_i$ if $i>j$. All those directed links constitute
a set $E'$, and $D(V,E')$ is an effective network. The proof is
straightforward and thus is omitted here. Effective network has a
clear hierarchical structure that the nodes closer to the source
have higher levels. The PSA with positive $\tau$ tends to put
high-degree nodes at higher levels. For a given undirected network
$G$, the extracted effective networks for different $\tau$ have
far different topologies, and we will see below that they exhibit
far different synchronization behaviors.

\begin{figure}
\scalebox{0.65}[0.65]{\includegraphics{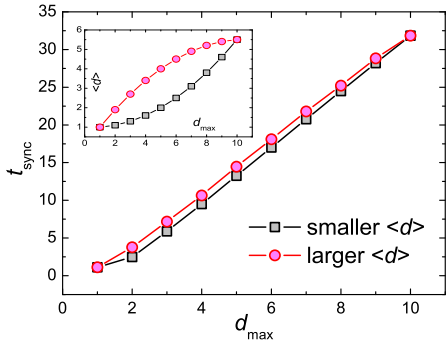}}
\caption{\label{fig:epsart} (Color online) The synchronization
time vs. the maximal depth. The black squares and red circles
represent two different configurations of the toy model and the
former has smaller average depth. The inset compares the average
depth of the two configurations. The network size is $N=1001$ and
the number of vertical links is $N_v=1000$. Each data point is
obtained by averaging over 1000 independent realizations.}
\end{figure}

We choose a benchmark individual dynamical system for the
numerical analysis, namely the \emph{Logistic map} \cite{May1976}.
For an arbitrary node $v_i$, we denote its state as $x_i$, which
evolves as $x_{i,t+1}=f(x_{i,t})-\varepsilon\sum_{j=1}^N
l_{ij}f(x_{j,t})$, where $\varepsilon$ is the overall coupling
strength, $l_{ij}$ is the element of the normalized Laplacian $L$,
$f(x)=ax(1-x)$ is a Logistic map with $a=4.0$ fixed in this paper.
According to Ref.\cite{Lind2004}, it can be proved that the
effective network will achieve the complete synchronization state
when the coupling strength satisfies the inequality
$1-e^{-\lambda}<\varepsilon<1$, where $\lambda$ is the Lyapunov
exponent of $f(x)$ at the current parameter.

To reveal how the topology affects the efficiency of
synchronization on effective networks, we study the
synchronization time versus the parameter $\tau$. The underlying
networks before PSA are generated by Barab\'asi-Albert (BA) model
\cite{Barabasi1999}. Given a $\tau$, we first extract the
effective network using the PSA, and then randomly assign each
node an initial state between 0 and 1, and then all the nodes
iterate according to the coupling equations for 2000 time steps,
which is long enough to drive the system to the synchronization
state. Then, we give each node a perturbation, that is, adding a
small number randomly chosen in (-0.005, 0.005). The
synchronization time, $t_{\texttt{sync}}$, is defined as the time
steps required for the system to return the synchronization state.
In the numerical implementation, we monitor the standard
deviation, $\sigma$, of the states of all nodes, and the system is
considered to be synchronized if $\sigma<10^{-6}$ and will not
increase any longer.

Figure 1 reports the synchronization time as a function of $\tau$.
Clearly, the synchronization time decreases remarkably as the
increasing of $\tau$. Although the effective networks
corresponding to different $\tau$ have exactly the same spectrum,
their synchronization behaviors are greatly different and the one
with higher-degree nodes in the topper positions synchronizes
faster (i.e., a larger $\tau$ leads to a faster synchronizing
process). In addition, there is a crossover of the curves for
different average degrees at $\tau_c\approx 0$ (represented by a
dot line in Fig. 1). Interestingly,   when $\tau$ is larger than $\tau_c$, the
network with smaller average degree will synchronize faster, in
contrast, the network with larger average degree will synchronize
faster. To understand these phenomena observed in Fig. 1, we next
study the structure properties of effective networks at different
$\tau$.

In an effective network, the number of links in the shortest path
from the source node to a node $v_i$ is called the \emph{depth} of
$v_i$, denoted by $d_i$, and the depth of the source node is zero.
The \emph{maximal depth} and the \emph{average depth} are defined
as $d_{\texttt{max}}=\texttt{max}\{d_{i}, 1\leq i \leq N\}$ and
$\langle d\rangle=\frac{1}{N}\sum^N_{i=1}d_i$, respectively. We
classify the links into two categories: vertical links are the
ones connecting two nodes with different depths (obviously, the
difference is one) while horizontal links connecting two nodes
with the same depth. Figure 2 reports the change of these
structural measurements with $\tau$. With the increasing of
$\tau$, the number of horizontal links will increase too, while
the maximal depth, the average depth and the number of vertical
links will decrease. Especially, the shapes of $d_\texttt{max}$
vs. $\tau$ and $\langle d\rangle$ vs. $\tau$ curves are very
similar to that of $t_{\texttt{sync}}$ vs. $\tau$ curve,
therefore one may guess that either $d_\texttt{max}$ or $\langle
d\rangle$ (or both) mainly determines the synchronization time. To
judge whether it is true, we propose a toy model, where $N+1$ nodes
are connected by $N$ vertical links starting from a single source node, resulting in a tree
structure. Given the maximal depth $d_{\texttt{max}}$ (in the
present simulations, we tune $d_{\texttt{max}}$ in the range
$1\leq d_{\texttt{max}} \leq 10$), we (i) put $\frac{N}{10}$
nodes in each of the $d_{\max}-1$ layers from $1\leq d \leq d_{\texttt{max}}-1$
and the remaining $N\left(
1-\frac{d_{\texttt{max}}-1}{10}\right)$ nodes in the
$d_{\texttt{max}}$-th layer; or (ii) put $\frac{N}{10}$ nodes in
each of the $d_{\max}-1$ layers  from $2 \leq d \leq d_{\texttt{max}}$ and the
remaining $N\left(1-\frac{d_{\texttt{max}}-1}{10}\right)$
nodes in the first layer. Each $d$-th layer node with $1\leq d
\leq d_{\texttt{max}}$ connects to (with an in-link) the node in
the $(d-1)$-th layer one by one. Clearly, if $d_{\texttt{max}}=1$
or $d_{\texttt{max}}=10$, the two methods lead to statistically
the same $\langle d\rangle$, while for $1< d_{\texttt{max}} <10$,
the former method (putting more nodes in the last layer) gives
longer $\langle d\rangle$. In the inset of Fig. 3, we report the
difference of average depths in the case of $N=1000$. It is
clearly observed in Fig. 3 that the synchronization time is
approximately proportional to the maximal depth, while almost not
affected by the average depth.

\begin{figure}
\begin{center}
\scalebox{0.8}[0.8]{\includegraphics{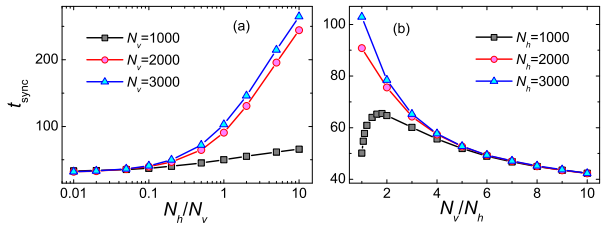}}
\end{center}
\caption{\label{fig:epsart} (Color online) Effects of (a) the
number of horizontal links and (b) the number of vertical links on
the synchronization time. The network size is $N=1001$ and each
data point is obtained by averaging over 1000 independent
realizations.}
\end{figure}

Further more, to reveal the roles of the vertical and horizonal
links, we randomly add some vertical and horizonal links in the
above-mentioned tree topologies. As shown in Fig. 4(a), the
horizontal links hinder the synchronization. In the case of
minimal number of vertical links (i.e., $N_v=1000$), the
horizontal links, together with vertical links, make up some
triangles in the form $\{v_i\rightarrow v_j, v_i\rightarrow v_k,
v_j\rightarrow v_k\}$ with $v_j$ and $v_k$ in the same layer; and
in the cases of abundant vertical links (i.e., $N_v=2000$ and
$N_v=3000$), they make up two types of triangles, respectively
with $v_j$ and $v_k$ in the same layer and with $v_i$ and $v_j$ in
the same layer. These triangles introduce additional disturbances
among individuals, and thus slow down the synchronization. The
simulations suggest a logarithmic relation between
$t_{\texttt{sync}}$ and $N_h$ at large $N_h$.

The effects of vertical links are twofold. Firstly,
more vertical links
are helpful for the propagation of controlling signals from the
upper-layer nodes to the lower-layer nodes.
At the same time,  the additional
vertical links will increase the number of triangles, and thus
hinder the synchronization.  These two competing
effects bring nontrivial  role of
vertical links. In simulations, we find that when there are not so many horizontal
links, adding vertical links will first hinder synchronization, and then
enhance synchronization, leading to a maximal synchronization time (Fig. 4(b)).
When  the
network possesses  abundant horizontal  links, the more vertical
links will accelerate the synchronization. Simulation results in
Fig. 4(a) and Fig. 4(b) suggest that when the network consists
abundant vertical and horizontal links, the ratio $N_h/N_v$ mainly
determines the synchronization time: the larger the ratio, the
slower the synchronizing process. This gives a qualitative
explanation of the crossover observed in Fig. 1. When $\tau>0$,
the maximal depths of networks with different average degrees are
more or less the same, but $N_h/N_v$ in the case of $\langle
k\rangle=10$ is much larger than the cases of $\langle k\rangle=4$
and $\langle k\rangle=6$. Therefore, when $\tau>0$, a network with
larger average degree synchronizes even slower than the one with
smaller average degree.

In conclusion, we have proposed a preferential selection algorithm
to extract effective networks from any given connected undirected
network. Although the spectra of all effective networks are the
same, their synchronization time can be far different. The key
ingredient affecting the synchronization time is the maximal depth
of an effective network: the larger the $d_{\texttt{max}}$, the
longer the $t_{\texttt{sync}}$. The secondary ingredient is the
number of links. In the cases of abundant links, the smaller
number of horizontal links and larger number of vertical links
will lead to a faster synchronization. Actually, with fixed
$d_{\texttt{max}}$, our simulation indicates that the ratio
$N_h/N_v$ mainly determines the synchronization time. Under the
framework of master stability analysis
\cite{Pecora1998,Barahona2002}, the majority of previous studies
on network synchronization only focus on the spectral analysis of
the coupling matrix \cite{Arenas,Chen}, and it has been shown that
synchronization time is proportional to $1/\lambda_2$  for typical
networks \cite{Almendral07}, but this is not the case for the
effective networks we propose, which have exactly the same
spectrum yet exhibit noticeably different synchronization
behaviors. Based on the results reported here, we claim that the
spectral analysis is helpful but not enough for the understanding
of network synchronization. Our findings not only emphasize again
the fundamental importance of directivity
\cite{Nishikawa2006,Nishikawa2007}, but also provide insights into
the roles of horizontal and vertical links in the synchronizing
process.

This work is partially supported by Hong Kong Baptist University.
 T.Z. and M.Z. acknowledge the
National Natural Science Foundation of China under Grant Nos.
10635040 and 10805045.

\end{document}